\documentclass[11pt]{article}
\usepackage{amsmath,amssymb,amsthm,mathrsfs,stackrel}
\usepackage{array}
\usepackage{graphicx,psfrag,epsf}
\usepackage{multirow}
\usepackage{multicol}
\usepackage{multirow}
\usepackage{booktabs}
\usepackage{natbib}
\usepackage{appendix}
\usepackage{float}
\usepackage[colorlinks,citecolor=blue,linkcolor=blue,urlcolor=blue]{hyperref}

\newtheorem{theorem}{Theorem}

\newtheorem{corollary}{Corollary}
\newtheorem{remark}{Remark}
\newtheorem{condition}{Condition}

\begin{document}

\title{A powerful transformation of quantitative responses for biobank-scale association studies}

\author{Yaowu Liu\thanks{Joint Lab of Data Science and Business Intelligence,
  Center of Statistical Research, Southwestern University of Finance and Economics, Chengdu, Sichuan, 611130, China.} \; and \; Tianying Wang\thanks{Department of Statistics, Colorado State University, Fort Collins, CO, 80523, USA.}}

\date{}

\maketitle
\thispagestyle{empty}
\baselineskip=20pt

\begin{abstract}
In linear regression models with non-Gaussian errors, transformations of the response variable are widely used in a broad range of applications. Motivated by various genetic association studies, transformation methods for hypothesis testing have received substantial interest. In recent years, the rise of biobank-scale genetic studies, which feature a vast number of participants that could be around half a million, spurred the need for new transformation methods that are both powerful for detecting weak genetic signals and computationally efficient for large-scale data. 
In this work, we propose a novel transformation method that leverages the information of the error density. This transformation leads to locally most powerful tests and therefore has strong power for detecting weak signals. To make the computation scalable to biobank-scale studies, we harnessed the nature of weak genetic signals and proposed a consistent and computationally efficient estimator of the transformation function. Through extensive simulations and a gene-based analysis of spirometry traits from the UK Biobank, we validate that our approach maintains stringent control over type I error rates and significantly enhances statistical power over existing methods.
\end{abstract}

\noindent
 \textbf{Keywords:} Kernel methods; Local alternative; Large sample size; Weak signals.

\pagestyle{plain}

\section{Introduction}\label{Sec:1}

Transformation of the response variable has a long-standing interest in statistics, especially when the residuals are non-Gaussian (e.g., skewed). Popular transformation methods can be traced back decades ago, including the well-known transformation by~\cite{box1964analysis}, the shifted power transformation~\citep{atkinson1982regression}, and the `transform-on-both-sides' approach that applies the same transformation to both the response and its mean function~\citep{carroll1984power}. The purposes of these transformations are mainly about estimation, e.g., constructing unbiased estimators or providing a better interpretation of the effect estimates. 

Over the past two decades, the burgeoning field of genetic association studies has spurred an increasing interest in transformation methods for hypothesis testing problems. 
This work focuses on gene-based (or variant set) association studies for a quantitative trait, where one is interested in testing the association between genetic variants within a gene or a set and a continuous trait. 
The conventional test statistics in variant set analysis are developed under the Gaussian linear regression models, such as the Burden test~\citep{li2008methods,price2010pooled} and the Sequence Kernel Association Test (SKAT)~\citep{wu2011rare}. However, for many complex traits, 
the residual distributions are markedly non-normal, indicating the violation of the normality assumption. Examples include the peak expiratory flow (PEF), whose residual is highly skewed even after the log transformation of the trait (see Section \ref{sec:data}). Consequently, the null distributions of these tests are only valid asymptotically, and in instances of small or modest sample sizes, these methodologies may struggle to preserve the nominal type I error~\citep{mccaw2020operating}. To overcome the type I error inflation posed by non-normal residuals in genetic association studies, the rank-based inverse normal transformation (INT) has emerged as a favored solution~\citep{beasley2009rank}. This method aligns the sample quantiles of the residuals with those of the standard normal distribution, enhancing the normality of residuals and thereby ensuring control of type I errors, even in studies with limited sample sizes. Further refinements in the application of INT have been proposed to boost the statistical power of these studies~\citep{sofer2019fully,mccaw2020operating}.

The importance of collecting large samples in genetic association studies has been widely acknowledged for increasing the statistical power of the detection of associated genetic variants~\citep{spencer2009designing}. With the advancement in next-generation sequencing technology and the decrease in sequencing cost, biobank-scale studies have become more and more prevalent in recent years~\citep{lu2022current,coppola2019biobanking}. These studies, featuring massive participant numbers, set a new trend and provide extensive insights into genomic and genetic variations. To name a few, the UK Biobank contains over half a million participants~\citep{sudlow2015uk}, and the All of Us Research Program aims to compile a comprehensive database from one million U.S. residents. Other large-scale studies that encompass a very broad segment of a population, such as BioBank Japan \citep{kanai2018genetic}, TOPMed \citep{taliun2021sequencing}, H3Africa \citep{consortium2014enabling}, Million Veteran Programme \citep{giri2019trans}, contribute valuable genetic and phenotypic data extensively to the scientific community. 

The massive sample sizes bring both benefits and challenges, and spur the need for new transformation methods that are suitable for biobank studies. First, the massive sample sizes enhance the accuracy of asymptotic approximations based on the central limit theorem. Therefore, the asymptotic null distributions could still lead to controlled type I error even if residuals deviate from normality substantially. The widely used transformation in genetic association studies, namely INT, which was developed a decade ago, mainly aims to protect the type I error and may not have the best possible power. For instance, the INT relies solely on the ranks of residuals and overlooks other information that potentially helps power improvement, such as the distributional shape of residuals. Hence, for biobank-scale studies, there is an increasing demand for transformation methods that could improve the testing power. Furthermore, it is well-recognized that many complex traits are often influenced by a large number of genetic variants, with most of them contributing subtly to the trait~\citep {wu2011rare}. The massive number of participants makes it possible to detect the genetic variants with weak effects. Therefore, in biobank studies, one would be particularly interested in boosting the power of detecting weak genetic signals. Last, while the massive sample size brings merits on the statistical side, it escalates the computational demands of analysis and requires methods that can be computed efficiently. 
We note that efficient computation is a common research topic for biobank-scale studies. Many efforts have been made to speed up the computation of various statistical methods~\citep[e.g.,][]{loh2015efficient,jiang2019resource}.

Based on these characteristics caused by the massive samples in biobank studies, our development of transformation methods has two goals: one goal is to overcome the statistical challenge of detecting weak genetic signals, and the other is 
to navigate the computational complexities inherent in analyzing large-scale data. In other words, the transformation methods should achieve the two goals simultaneously in order to be practically useful. These two goals also reflect broader themes in large-scale data analysis, where statistical and computational challenges often intersect~\citep{donoho2015higher}.% \yl{This reference seems not for this sentence.} 

%We tackle the two major challenges, namely weak signals and computational burden, in a synergistic way, wherein the weak signals are harnessed to mitigate the computational burden. 

To tackle the statistical challenge (i.e., enhancing the power of detecting weak signals),
we propose a transformation function that leverages the error distribution. We show that the proposed transformation yields tests that are locally most powerful under the forms of Burden and SKAT test statistics, and therefore maintain strong power under weak signals. The proposed optimal transformation depends on the unknown error distribution and therefore needs to be estimated. The massive sample size in biobank-scale studies leads to a heavy computational burden for the estimation of the transformation. We leverage the characteristic of weak signals, which is a challenge on the statistical side, to propose a consistent estimator of the transformation that is also computationally scalable for biobank-scale studies. We show that the test with estimated transformation has asymptotically equivalent power to the locally optimal tests. Our methodology highlights the novel strategy that the statistical and computational challenges are not tackled separately, but in a synergistic way where the weak signals (i.e., the statistical challenge) are harnessed to mitigate the computational burden.

The rest of the paper is organized as follows. In Section~\ref{Sec:background}, we introduce the background that includes the regression models and different tests in genetic association studies. In Section~\ref{Sec:method}, we propose a transformation that achieves local optimal power and establishes the theoretical properties of the test statistics based on the proposed transformation. In Section~\ref{sec:simu}-\ref{sec:data}, we conduct extensive numerical experiments to evaluate the performance of the proposed transformation and demonstrate its superiority using UK Biobank data. All the technical proofs, additional simulation results, and real data analysis results are provided in the supplementary materials.

\section{Background}\label{Sec:background}

\subsection{The model and notations}\label{Sec:2:1}

Let $y_i$ denote the response variable, $\mathbf{z}_i$ denote a $q \times 1$ vector of covariates that also include the intercept term, and $\mathbf{g}_i$ denote a $p \times 1$ vector of explanatory variables for the $i$-th subject, where $i = 1,2,\cdots,n$.
We consider a linear regression model
\begin{equation}\label{Eq:model}
  y_i = \mathbf{z}_i^\top\boldsymbol{\alpha} + \mathbf{g}_i^\top\boldsymbol{\beta} + \epsilon_i,
\end{equation}
where $\boldsymbol{\alpha}$ is a $q \times 1$ vector of coefficients, $\boldsymbol{\beta}$ is a $p \times 1$ vector of coefficients, and the error terms $\epsilon_i$'s are independently and identically distributed (i.i.d.) continuous variables with a density function $f(\cdot)$ that has $E(\epsilon_i) = 0$ and $\text{Var}(\epsilon_i) = \sigma^2 < +\infty$. The question of interest is to test whether the explanatory variables $\mathbf{g}_i$ have any effects on the response variable $y_i$ conditional on the covariates $\mathbf{z}_i$, i.e.,
\begin{equation}\label{Eq:hypo}
 H_0: \boldsymbol{\beta} = 0 \quad\quad \text{against} \quad\quad H_a: \boldsymbol{\beta} \neq 0.
\end{equation}

In gene-based association analysis, $y_i$ is the quantitative trait of interest, e.g., height, lipid levels and spirometry measurement. The covariates $\mathbf{z}_i$ could include the intercept term, age, gender, and variables that control for population stratification. Explanatory variables are genetic variants such as Single Nucleotide Polymorphisms (SNPs), and $\mathbf{g}_i$ is the genotypes of $p$ SNPs in a gene (or set), with each element often coded as 0, 1, and 2 to represent the number of minor alleles. In gene-based association studies,  $p$ is typically at most a few hundred and remains much smaller than the sample size  $n$, which can reach up to half a million in large-scale studies such as the UK Biobank. Hence, the fixed $p$ setting is a well-accepted for gene-based (or variant set) analysis in the literature~\citep[e.g.,][]{liu2022minimax, wang2022integrated, li2022simultaneous}.  The hypothesis~\eqref{Eq:hypo} is to investigate whether any of the SNPs in a gene is associated with the trait conditional on the covariates. Typically, tens of thousands of genes (or variant sets) across the genome are examined, and the main goal is to identify the associated genes (or variant sets). Statistically, this means that one needs to perform the test for~\eqref{Eq:hypo} repeatedly a vast number of times. 

Given our main model (Eq~\eqref{Eq:model}) and hypothesis (Eq~\eqref{Eq:hypo}), we introduce additional notations and conditions for subsequent sections.
 Let $\mathbf{Y}$ be an $n \times 1$ vector of responses whose $i$-th element is $y_i$, $\mathbf{Z}$ be an $n\times q$ matrix of covariates whose $i$-th row is $\mathbf{z}_i$, and $\mathbf{G}$ be an $n\times p$ matrix whose $i$-th row is $\mathbf{g}_i$. Denote $\mathbf{P}_Z = \mathbf{I}_n - \mathbf{Z}(\mathbf{Z}^\top\mathbf{Z})^{-1}\mathbf{Z}^\top$ as the projection matrix of $\mathbf{Z}$, where $\mathbf{I}_n$ is an $n\times n$ identity matrix. Let $\tilde{\mathbf{g}}_i $ be the $i$-th row of $\tilde{\mathbf{G}} = \mathbf{P}_Z\mathbf{G}$, i.e., $\tilde{\mathbf{g}}_i$ is the residual of $\mathbf{g}_i$ after projecting on the linear space spanned by $\mathbf{Z}$. Let 
\begin{equation}\label{eq:alpha:est}
  \hat{\boldsymbol{\alpha}} = (\mathbf{Z}^\top\mathbf{Z})^{-1}\mathbf{Z}^\top \mathbf{Y}.
\end{equation}
denote the least square estimator of $\boldsymbol{\alpha}$ under the null $H_0$.
Last, we introduce a regularity condition on the design matrix that will be used in the asymptotic theories. 

\begin{condition}\label{Con:1}
  Let $\mathbf{A} = (a_{ij})$ be an $n\times p$ design matrix.  The design matrix satisfies (1) $\lim_{n\rightarrow +\infty} n^{-1}\mathbf{A}^\top\mathbf{A} = \boldsymbol{\Sigma}$, where $\boldsymbol{\Sigma}$ is a $p\times p$ positive definite matrix; 
  
  \noindent (2) $\lim_{n\rightarrow +\infty} \max_{1\leq i \leq n} \mathbf{a}_i^\top(\mathbf{A}^\top\mathbf{A})^{-1}\mathbf{a}_i \rightarrow 0$, where $\mathbf{a}_i$ is the $i$-th row of $\mathbf{A}$.
\end{condition}
This is a common and standard condition for asymptotic theories in regression models~\citep[see, e.g.,][]{van2000asymptotic}. With all the necessary notations and conditions, we will now present the general forms of the two widely used test statistics, setting the stage for the introduction of the local most powerful tests in Section~\ref{Sec:method}.

\subsection{Test statistics}\label{Sec:2:2}

In genetic association studies, two types of test statistics are commonly used for testing the global null~\eqref{Eq:hypo} in practice. These test statistics are derived under the Gaussian regression setting, and the residuals (or responses) are not transformed. Our proposed methods, which apply transformations on the residuals, include the existing tests without transformations as special cases by taking the transformation to be the identity function. 
To keep the notation concise, we directly present the general forms of these test statistics with transformations. The motivation for considering the general forms will be discussed in Section~\ref{Sec:method}.

Consider a generic transformation function $\psi(\cdot)$ that is continuously differentiable and satisfies $\sigma^2_\psi = var\{\psi(\epsilon_i)\} < +\infty$. 
Let $\mathbf{Y}_\psi$ be an $n\times 1$ vector whose $i$-th element is $\psi(y_i - \mathbf{z}_i^\top\hat{\boldsymbol{\alpha}})$, i.e., $\mathbf{Y}_\psi$ is the vector of transformed residuals. We estimate $\sigma^2_\psi$ by $\hat{\sigma}_\psi^2 = \mathbf{Y}_\psi^\top\mathbf{P}_Z\mathbf{Y}_\psi/(n-q)$, where $\mathbf{P}_Z$ is defined in Section~\ref{Sec:2:1}. 

In genetic variant set association studies, it has been observed that the genetic variants in a set could be all protective or harmful for a disease or trait~\citep{sham2014statistical}, i.e., the non-zero elements of $\boldsymbol{\beta}$ have the same sign under the alternative. To be powerful in this scenario, the linear type of tests is commonly employed. The general form of the linear test statistics with transformed residuals is given by
\begin{equation}\label{eq:linear:type}
     T_{\text{Linear},\psi} = \frac{1}{\sqrt{n}\hat{\sigma}_\psi}\sum_{i=1}^n   (\tilde{\mathbf{g}}_i^\top \mathbf{w}) \cdot \psi(y_i - \mathbf{z}_i^\top\hat{\boldsymbol{\alpha}}),
\end{equation}
where $\mathbf{w}$ is a vector of pre-specified weights. Large values of $|T_{\text{Linear},\psi}|$ indicate evidence against the null $H_0$. Let $T_{\text{Linear},I}$ denote the statistic $ T_{\text{Linear},\psi}$ with $\psi(\cdot)$ being the identity function (i.e., no transformation is performed).
When $\mathbf{w} = \mathbf{1}_p$, the linear test $T_{\text{Linear},I}$ reduces to the Burden test, which has become a popular method in genetic variant set association analysis~\citep{li2008methods,price2010pooled}. The Burden test is particularly powerful in the scenario with the same effect sign.

When the effects have different signs, the Burden test could significantly lose power since positive and negative effects could be canceled out in the linear statistic. In this situation, it is preferred to use SKAT, which is a quadratic test, and therefore, its power is insensitive to the effect signs. SKAT is derived from an empirical Bayes perspective and is a score test under a random effect setting~\citep{wu2011rare}.
Specifically, SKAT imposes a working assumption that the elements of $\boldsymbol{\beta}$ are independent and follow an arbitrary distribution with mean $0$ and variance $\tau^2$. Then, testing~\eqref{Eq:hypo} is equivalent to test the variance component $H_0: \tau^2 = 0$ against $H_a: \tau^2 > 0$. The SKAT statistic is derived based on the optimal test in the limiting situation that $\tau \rightarrow 0$ and, therefore, has the locally optimal power. The SKAT statistic with transformed residuals is
\begin{equation}\label{eq:SKAT:type}
   T_{\text{SKAT},\psi} = \frac{1}{n\hat{\sigma}_\psi^2}\sum_{i=1}^n \sum_{j=1}^n (\tilde{\mathbf{g}}_i^\top\tilde{\mathbf{g}}_j) \cdot \psi(y_i - \mathbf{z}_i^\top\hat{\boldsymbol{\alpha}})\psi(y_j - \mathbf{z}_j^\top\hat{\boldsymbol{\alpha}}).
\end{equation}
When $\psi(\cdot)$ is the identity function, then $T_{\text{SKAT},\psi}$ degenerates to the original SKAT statistic, which is denoted by $T_{\text{SKAT},I}$.

When the error distribution is Gaussian,
under the null $H_0$, $T_{\text{Linear},I}$ follows a normal distribution, while $T_{\text{SKAT},I}$ follows a mixture of chi-squared distributions~\citep{wu2011rare}. In the presence of non-Gaussian errors, the null distributions of $T_{\text{Linear},I}$ and $T_{\text{SKAT},I}$ are only valid asymptotically. Hence, in this situation, a direct application of the two tests could suffer from considerable type I error inflation when the sample size is not large.
To protect the type I error under non-Gaussian errors, the INT has become a popular method in genetic association studies~\citep{beasley2009rank,mccaw2020operating}. The INT increases the normality of the residuals and, therefore, speeds up the convergence of the null distributions of the test statistics.

The primary purpose of using INT in genetic association studies is to control type I errors. With the advent of biobank-scale projects such as the UK biobank data, since the sample size is extremely large (e.g., $n > 100,000$), the approximations based on the asymptotic null distributions could still be accurate even if the residuals (or responses) deviate the normal distributions substantially, especially for analyzing common variants in gene-based analysis. 
In other words, the large sample size alleviates the concern of the type I error inflation. This motivates us to pursue transformation methods that aim to improve the testing power in the presence of non-Gaussian errors. 

It is clear that the INT may not have the optimal power under non-Gaussian distributions since it only utilizes residual ranks, ignoring other information like distributional shape. For instance, consider two residual sets: one symmetric, the other highly skewed. If the residual ranks in the two sets are the same, then the transformed residuals would also be the same in the two sets. However, it can be seen intuitively that the optimal tests should depend on the density function of the residuals.

\section{The proposed method}\label{Sec:method}

\subsection{Why focus on weak signals?}
For a specific application such as genetic association studies, the task of practitioners is to determine which methods (i.e., the statistical test and transformation in our case) should be employed in the analysis. For strong and moderate signals, with a large sample size, the choice of testing methods is less critical since all the methods would have a high power. For instance, for a strong signal, suppose that the p-value of a test is $10^{-15}$ while the p-value of another test is $10^{-12}$.
Although one test provides stronger evidence than the other, both tests pass the significance threshold and identify this strong signal. Hence, for biobank-scale projects, it is of particular interest to detect weak genetic signals. In fact, the prevalence of weak genetic signals for many complex traits is one of the reasons for the launch of biobank-scale studies.

\subsection{The locally most powerful tests}
For the regression model~\eqref{Eq:model}, 
weak signals occur when the regression coefficients $\boldsymbol{\beta}$ have small magnitudes under the alternative hypothesis, a scenario known as local alternatives. To detect these weak signals effectively, we focus on tests that are powerful against local alternatives. Locally most powerful tests are particularly effective, as they maximize power for detecting small deviations from the null, making them well-suited for identifying weak genetic signals. This motivates us to derive the locally most powerful tests for the regression model~\eqref{Eq:model} with non-Gaussian errors.
In the Gaussian regression models, the optimality of the linear (or Burden) test is established under the fixed effect setting, while the optimality of SKAT is in the random effect setting that takes an empirical Bayes perspective~\citep{liu2022minimax}. Similarly, we consider both settings when deriving the locally most power tests. For now, we assume the nuisance parameter $\boldsymbol{\alpha}$ and the error density $f(\cdot)$ are known. 

\begin{theorem}\label{Thm:optimal:test}
  Suppose that $f(\cdot)$ is twice continuous differentiable on its support $(-\infty,+\infty)$.
  
  (a) (Fixed effects) Suppose that $\boldsymbol{\beta}$ is a vector of fixed effects. Denote $\mathbf{w}_\beta = \boldsymbol{\beta}/||\boldsymbol{\beta}||$ and $\tau = || \boldsymbol{\beta} ||$. Then, when $\tau\rightarrow 0$, the test statistic with optimal local power is
  \begin{equation}\label{Eq:linear:f}
     T_{\text{Linear},f} = \frac{1}{\sqrt{n}}\sum_{i=1}^n   (-\mathbf{g}_i^\top \mathbf{w}_\beta) \cdot \frac{f'(y_i - \mathbf{z}_i^\top\boldsymbol{\alpha} )}{f(y_i - \mathbf{z}_i^\top\boldsymbol{\alpha})}.
  \end{equation}

  (b)(Random effects) Suppose that $\boldsymbol{\beta}$ is a vector of random effects. Assume that $E(\boldsymbol{\beta}) = 0$ and $E(\boldsymbol{\beta}\boldsymbol{\beta}^\top) = \tau^2\mathbf{I}$, where $\mathbf{I}$ is an identity matrix and $\tau^2$ is a variance parameter. Then, when $\tau\rightarrow 0$, the test statistic with optimal local power is
  \[
    T_{Quad,f} = \frac{1}{n} \sum_{i=1}^n (\mathbf{g}_i^\top\mathbf{g}_i) \cdot \frac{f''(y_i - \mathbf{z}_i^\top\boldsymbol{\alpha})}{f(y_i - \mathbf{z}_i^\top\boldsymbol{\alpha})} +  \frac{1}{n}\sum_{i=1}^n \sum_{j\neq i}^n (\mathbf{g}_i^\top\mathbf{g}_j) \cdot \frac{f'(y_i - \mathbf{z}_i^\top\boldsymbol{\alpha})}{f(y_i - \mathbf{z}_i^\top\boldsymbol{\alpha})} \cdot \frac{f'(y_j - \mathbf{z}_j^\top\boldsymbol{\alpha})}{f(y_j - \mathbf{z}_j^\top\boldsymbol{\alpha})}.
  \]
\end{theorem}

The proof is given in the supplementary materials. We first consider the fixed effect setting. From Theorem~\ref{Thm:optimal:test} (a), it can be seen that the optimal test $T_{\text{Linear},f}$ has the same form as the linear test statistic in the Gaussian regression models (i.e., $T_{\text{Linear},I}$), expect that the residual $y_i - \mathbf{z}_i^\top\boldsymbol{\alpha}$ is replaced by $\frac{f'(y_i - \mathbf{z}_i^\top\boldsymbol{\alpha} )}{f(y_i - \mathbf{z}_i^\top\boldsymbol{\alpha})}$, which can be viewed as a transformation of the residual by the function $f'(\cdot)/f(\cdot)$. In other words, to account for possibly non-Gaussian errors, the error density is harnessed in the transformation of the residuals to improve the testing power. 

We next consider the random effect setting. The optimal test statistic $T_{\rm Quad,f}$ has a slightly different form compared to its counterpart in the Gaussian regression models. But as will be shown in Section~\ref{Secsub:alternative}, when detecting weak signals, $T_{\text{Quad},f}$ is asymptotically equivalent to 
\begin{equation}\label{Eq:SKAT:f}
    T_{\text{SKAT},f} = \frac{1}{n}\sum_{i=1}^n \sum_{j=1}^n (\mathbf{g}_i^\top\mathbf{g}_j) \cdot \frac{f'(y_i - \mathbf{z}_i^\top\boldsymbol{\alpha})}{f(y_i - \mathbf{z}_i^\top\boldsymbol{\alpha})} \cdot \frac{f'(y_j - \mathbf{z}_j^\top\boldsymbol{\alpha})}{f(y_j - \mathbf{z}_j^\top\boldsymbol{\alpha})},
\end{equation} 
since the interaction terms (i.e., $i\neq j$) in $T_{\text{Quad},f}$ are dominating asymptotically. 
Then, similar to the comparison between $T_{\text{Linear},f}$ and its counterpart $T_{\text{Linear},I}$ in the Gaussian regression models, $T_{\text{SKAT},f}$ replaces the residual in the SKAT statistic, namely $T_{\text{SKAT}, I}$, by the transformed residuals. Hereafter, we will consider the SKAT-type test $T_{\text{SKAT},f}$.

In summary, Theorem~\ref{Thm:optimal:test} indicates that from Gaussian errors to non-Gaussian errors, we can keep the form of test statistics and only need to make suitable transformations on the residuals in order to have the locally optimal power. 
In fact, the locally optimal tests $T_{\text{Linear},f}$ and $T_{\text{SKAT},f}$ include their counterparts in the Gaussian regression models as special cases. When $f(\cdot)$ is the density of the standard Gaussian distribution, it is easy to see that $f'(\cdot)/f(\cdot)$ is the identity function. Then, $T_{\text{Linear},f}$ and $T_{\text{SKAT},f}$ were reduced to their counterparts in the Gaussian regression model, respectively. Hence, it is clear that we always need to transform the residuals. There is no transformation for Gaussian errors because the transformation happens to be the identity function.

 The transformation $f'(\cdot)/f(\cdot)$ is the logarithmic derivative of $f(\cdot)$, which aligns with the form of a score function in likelihood-based inference. Score functions are fundamental in constructing locally most powerful tests \citep{lin1997variance}, which capture the sensitivity of the likelihood to parameter changes. Since our proposed test in Theorem \ref{Thm:optimal:test} is locally most powerful, using the transformation in this form naturally enhances the efficiency and interpretability of our method. 

\subsection{The proposed test statistics}

When the nuisance parameter $\boldsymbol{\alpha}$ is unknown, we replace $\boldsymbol{\alpha}$ in the test statistics $T_{\text{Linear},f}$ and $T_{\text{SKAT},f}$ by the least square estimator $\hat{\boldsymbol{\alpha}}$ defined in~\eqref{eq:alpha:est}, which is a consistent estimator under the null~\citep{sen1994large}. 
Further, we use $\tilde{\mathbf{g}}_i$ instead of $\mathbf{g}_i$, where $\tilde{\mathbf{g}}_i$ is defined in Section \ref{Sec:2:1} and represents the residual of $\mathbf{g}_i$ after projecting onto the linear space spanned by $\mathbf{Z}$. This adjustment is crucial for several reasons, which will be discussed later in Remark \ref{remark:null}. Then, we obtain the linear-type statistic $T_{\text{Linear},\psi}$ in~\eqref{eq:linear:type} and SKAT-type statistic $T_{\text{SKAT},\psi}$ in~\eqref{eq:SKAT:type} introduced in Section~\ref{Sec:2:2}.
To have the locally optimal power, we choose the transformation $\psi(\cdot) = f'(\cdot)/f(\cdot)$. 

Our proposed test statistics $T_{\text{Linear},\psi}$ and $T_{\text{SKAT},\psi}$ can also be viewed from another perspective. Note that the original linear and SKAT test statistic under the Gaussian regressions can be simplified, e.g., 
\[
T_{\text{Linear},I} = 1/(\sqrt{n}\hat{\sigma}_\psi)\sum_{i=1}^n  (\mathbf{w}^\top\tilde{\mathbf{g}}_i)(y_i - \mathbf{z}_i^\top\hat{\boldsymbol{\alpha}}) = 1/(\sqrt{n}\hat{\sigma}_\psi)\sum_{i=1}^n  (\mathbf{w}^\top\tilde{\mathbf{g}}_i)y_{i},
\]
 where $\hat\sigma^2_\psi$ is the estimated error variance after transformation and defined in Section \ref{Sec:2:2}.
Comparing this form of $T_{\text{Linear},I}$ with $T_{\text{Linear},\psi}$ in~\eqref{eq:linear:type},
it can be seen that our method can be viewed as the following operationally: treating the $\psi$-transformed residual as the response $y_i$ in model~\eqref{Eq:model}, and then performing the testing procedure in the Gaussian regression models as usual. In other words, compared to the existing tests in Gaussian regressions, our method only replaces $y_i$ by the transformed residuals, which provides a simple interpretation of our methodology in practice. 

 The error density $f(\cdot)$ is often unknown and needs to be estimated. In Section~\ref{sec:est}, we will discuss the important issue of estimating the optimal transformation $f'(\cdot)/f(\cdot)$. Before that, we establish the asymptotic null distributions of $T_{\text{Linear},\psi}$ and  $T_{\text{SKAT},\psi}$.

\subsection{Asymptotic null distributions}

While our proposed transformation is $f'(\cdot)/f(\cdot)$, we establish the asymptotic null distributions of $T_{\text{Linear},\psi}$ and $T_{\text{SKAT},\psi}$ under a general transformation $\psi(\cdot)$ when the sample size $n$ goes to $+\infty$. The proof of the following theorem is provided in the supplementary materials.

\begin{theorem}\label{Thm:null}
  Suppose that the design matrix $\tilde{\mathbf{G}}$ satisfies the regularity Condition~\ref{Con:1} and let $\boldsymbol{\Sigma} = \lim_{n\rightarrow +\infty} n^{-1}\tilde{\mathbf{G}}^\top\tilde{\mathbf{G}}$. Assume that $var\{\psi(\epsilon_i)\} < +\infty$ and $E|\psi'(\epsilon_i)| < +\infty$. Then, under $H_0$, we have (a) $T_{\text{Linear},\psi} \stackrel[]{d}{\rightarrow} N(0, \mathbf{w}^\top\boldsymbol{\Sigma}\mathbf{w})$; (b) $T_{\text{SKAT},\psi} \stackrel[]{d}{\rightarrow} \sum_{j=1}^p \lambda_j \chi^2_{1,j}$, where $\chi^2_{1,j}$'s denote independent $\chi^2_{1}$ variables and $\lambda_j$'s are the eigenvalues of $\boldsymbol{\Sigma}$.
\end{theorem}
Theorem~\ref{Thm:null} indicates that the asymptotic null distributions of $T_{\text{Linear},\psi}$ and $T_{\text{SKAT},\psi}$ are the same as those of their counterparts, namely $T_{\text{Linear},I}$ and $T_{\text{SKAT},I}$, in the Gaussian regression, respectively. In fact, the asymptotic null distributions do not depend on the transformation function $\psi(\cdot)$. This implies that the type I error is still protected asymptotically even if the optimal transformation function $f'(\cdot)/f(\cdot)$ is not used. In other words, the choice of the transformation $\psi(\cdot)$ only affects the power but not the type I error control in asymptotics. As we highlighted before, the immense sample sizes in biobank-scale data help improve the accuracy of type I errors based on the asymptotic null distributions in finite samples.  

%Under the null $H_0$, $\hat{\boldsymbol{\alpha}}$ is a $\sqrt{n}$-consistent estimator for the true value $\boldsymbol{\alpha}$~\citep{sen1994large}. 

\begin{remark}\label{remark:null}
To make the asymptotic null distributions in Theorem~\ref{Thm:null} hold, it is important that $\tilde{\mathbf{g}}_i$ (i.e., the residual of $\mathbf{g}_i$ regressing on $\mathbf{Z}$) is used in the test statistics rather than $\mathbf{g}_i$. In other words, covariates are adjusted again after transformation. We note that double adjustments for covariates were also suggested for the INT method~\citep{sofer2019fully}.
\end{remark}

\begin{remark}\label{remark:MAF}
In genetic association studies, the accuracy of the asymptotic approximation of the null distribution in Theorem~\ref{Thm:null} also depends on the minor allele frequency (MAF) of SNP. In general, there is an inverse relationship between MAF and the required sample size for achieving an accurate asymptotic approximation. Our this work focuses on the analysis of common variants, defined as those with an MAF $>$ 5\%. It is important to note that our transformation approach is applicable in situations where the asymptotic null distribution can effectively control the type I error rate. Therefore, our method may be suitable for analyzing rare variants with MAF $<$ 5\% when the sample size is sufficiently large, such as when $n > 10,000$. Investigating the practical performance of our approach in rare-variant analysis is an interesting direction for future research.
\end{remark}

\subsection{Asymptotic setting of weak signals and an equivalent test statistic}\label{Secsub:alternative}

 In the asymptotic setting, the alternative should be neither completely separable nor completely inseparable from the null. Otherwise, the choice of testing methods becomes meaningless since different methods would have the same result. For instance, if the nonzero $\boldsymbol{\beta}$ is fixed under the alternative, the power of any reasonable tests would converge to 1 as the sample size $n\rightarrow +\infty$ because the null and alternative are completely separable asymptotically. Since the regression model~\eqref{Eq:model} is a parametric model, the convergence rate of the effects should be $||\boldsymbol{\beta}|| = O(1/\sqrt{n})$ in order to have a nontrivial asymptotic power~\citep{van2000asymptotic}, i.e., the asymptotic power is neither 1 nor the significance level. In summary, our asymptotic setting of weak signals means the local alternatives where $||\boldsymbol{\beta}|| = O(1/\sqrt{n})$. We will see next that the situation of weak signals benefits the methodology development.

The theorem below shows that $T_{\text{Quad},f}$ and $T_{\text{SKAT},f}$ defined in Theorem~\ref{Thm:null} and~\eqref{Eq:SKAT:f} are asymptotically equivalent under the local alternatives.
\begin{theorem}\label{Thm:equivalence}
  Under both the null and the local alternative that $||\boldsymbol{\beta}|| = O(1/\sqrt{n})$, $T_{\text{Quad},f} - T_{\text{SKAT},f} \rightarrow c_0$, where $c_0$ is a constant. Hence, the two test statistics are asymptotically equivalent.
\end{theorem}
We highlight that it is more convenient to use $T_{\text{SKAT},f}$ in practice for the following reasons. The asymptotic null distribution of $T_{\text{Quad},f}$ involves the constant $c_0$ in Theorem~\ref{Thm:equivalence}, which depends on $f(\cdot)$ or the transformation and therefore needs to be estimated. In contrast, as indicated by Theorem~\ref{Thm:null}, the asymptotic null distribution of $T_{\text{SKAT},f}$ is a mixture of chi-squared distribution and does not depend on the transformation.
Note that equivalence in Theorem~\ref{Thm:equivalence} is established under local alternatives.  This implies that in scenarios with weak signals, we have the advantage of employing the more convenient $T_{\text{SKAT},f}$.

\subsection{A consistent estimator of the optimal transformation}\label{sec:est}
Next, we discuss the estimation of the optimal transformation $f'(\cdot)/f(\cdot)$. The feature of weak signals helps to address the computational challenges of estimating the transformation. Since the optimal transformation is determined by the error density $f(\cdot)$, an estimator of $f(\cdot)$ can directly lead to an estimator of $f'(\cdot)/f(\cdot)$. It is common to use the kernel method for density estimation. Here, we propose a kernel estimator for $f(\cdot)$, and the Gaussian kernel is used. Let $K(\cdot)$ denote the density of the standard Gaussian distribution. With $h_n$ as the bandwidth, our density estimator is given by
\begin{equation}\label{Eq:kernel}
  \hat{f}_n(x) = \frac{1}{n} \sum_{i=1}^n \frac{1}{h_n} K\left(\frac{y_i - \mathbf{z}_i^\top\hat{\boldsymbol{\alpha}} - x }{h_n}\right).
\end{equation}

In Eq~\eqref{Eq:kernel}, the residuals under the null (i.e., $y_i - \mathbf{z}_i^\top\hat{\boldsymbol{\alpha}}$) are used in kernel density estimation. Since the purpose of transformation is to increase the power when $H_a$ is true, the consistency of $\hat{f}_n(x)$ should be studied under the alternative instead of under the null. Note that $\epsilon_i = y_i - \mathbf{z}_i^\top\boldsymbol{\alpha} - \mathbf{g}_i^\top\boldsymbol{\beta}$. In general, a consistent kernel estimator would require residuals under the alternative, i.e., $y_i - \mathbf{z}_i^\top\hat{\boldsymbol{\alpha}} - \mathbf{g}_i^\top\hat{\boldsymbol{\beta}}$, where $\hat{\boldsymbol{\beta}}$ is some estimator of $\boldsymbol{\beta}$. In gene-based (or variant set) analysis, if residuals under the alternative are used, then density estimation needs to be performed for every gene (or variant set) since $\boldsymbol{\beta}$ is estimated every time. When the sample size $n$ is at the biobank-scale (e.g., $n\approx 220,000$ in our real data analysis), density estimation is not fast and could take several minutes (see section~\ref{Secsub:comput:time} about the computation time). In gene-based analysis, hundreds of thousands of genes are scanned across the genome. Hence, it is computationally intractable to run density estimation for every gene. In contrast, the residuals under the null are shared by all the genes. The estimator~\eqref{Eq:kernel} only needs to be calculated once in gene-based analysis and, therefore, is computationally feasible. This computational advantage is similar to that of SKAT (see remark~\ref{remark:2} for details), which plays a critical role in the wide use of SKAT in practice.

In addition to computational efficiency, it is crucial to recognize that $\hat{f}_n(x)$ defined in~\eqref{Eq:kernel} is a consistent estimator for the error density $f(\cdot)$.  Without this consistency, the test with the estimated transformation would not achieve optimal power. However, the consistency of $\hat{f}_n(x)$ does not generally hold. Fortunately, it is maintained under weak signals with $||\boldsymbol{\beta}|| = O(1/\sqrt{n})$. The intuition is that the bias caused by the residuals under the null is asymptotically negligible under weak signals. 
Hence, the computationally efficient estimator $\hat{f}_n(x)$ benefits from the characteristic of weak signals, which is a statistical challenge, to achieve consistency.   

Furthermore, under some mild conditions, the derivative of the kernel density estimator consistently estimates the derivative of the density~\citep{silverman1978weak}. Therefore, we use the derivative of $\hat{f}_n(x)$, denoted by $\hat{f}_n'(x)$, to estimate $f'(x)$. Finally, the transformation function $f'(\cdot)/f(\cdot)$ is estimated by $\hat{f}'_n(x)/\hat{f}_n(x)$. The following theorem indicates that the linear-type and SKAT-type tests with the estimated transformation $\hat{f}'_n(x)/\hat{f}_n(x)$ also have the optimal local power asymptotically. The proof is given in the supplementary materials.

\begin{theorem}\label{Thm:kernel}
  Suppose that $f(\cdot)$ has uniformly continuous second derivative and 
 $E[f'(\epsilon_i)/f(\epsilon_i)]^2 < +\infty$. The design matrix $\tilde{\mathbf{G}}$ satisfies the regularity Condition~\ref{Con:1}.
  The bandwidth in the kernel estimator $f_n(\cdot)$ is $h_n = n^{-1/5}$.
  Under the local alternatives that $||\boldsymbol{\beta}|| = O(1/\sqrt{n})$, the linear test $T_{\text{Linear},\psi}$ (or the quadratic test $T_{\text{SKAT},\psi}$) with $\psi(\cdot) = f'(\cdot)/f(\cdot)$ has the same asymptotic power as that with $\psi(\cdot) = \hat{f}'_n(\cdot)/\hat{f}_n(\cdot)$.
\end{theorem}
 
Last, the asymptotic null distributions also hold with the estimated transformation function, which is presented in the corollary below.
\begin{corollary}
  Under the same conditions of Theorem~\ref{Thm:kernel}, the results in Theorem~\ref{Thm:null} also hold with $\psi(\cdot) = \hat{f}'_n(\cdot)/\hat{f}_n(\cdot)$.
\end{corollary}

\begin{remark}\label{remark:2}
  In gene-based (or variant set) analysis, a testing method needs to be performed repeatedly a large number of times. The reason that our method is computationally efficient is similar to that of SKAT, i.e., the time-consuming part of the testing procedure only needs to be performed once for all the genes or variant sets. The first step in SKAT is model fitting, which is computationally intensive. Since SKAT only requires fitting the null model that is shared by all the variant sets rather than the models under the alternative, modeling fitting only needs to be carried out once for SKAT. Similarly, in our method, while estimating the transformation function is computationally expensive when the sample size is large, it also only needs to be done once since we use the residuals under the null in the estimation instead of the residuals under the alternatives. 
\end{remark}

\begin{remark}\label{remark:3}
In Theorem~\ref{Thm:kernel}, the non-parametric estimation of the transformation function affects the non-central parameter and, therefore, the power of the test, but not the asymptotic normality of the test statistic. As stated in Theorem~\ref{Thm:null}, the asymptotic null distribution holds for a broad class of transformation functions, implying the transformation function acts as a nuisance parameter with respect to asymptotic normality. Hence, the slower convergence rate of the non-parametric estimation does not comprise the $\sqrt{n}$ convergence rate of asymptotic normality. 
\end{remark}

\subsection{Extensions}\label{sec:ext}

We have discussed applying the transformation $\hat{f}'_n(\cdot)/\hat{f}_n(\cdot)$ to the linear-type and SKAT-type tests. As a transformation method, it can also be directly applied to other tests. For instance, \cite{liu2022minimax} recently proposed another quadratic-form test, which is referred to as the Minimax Optimal Ridge-type Set Test (MORST), to have robust power in variant set analysis. Since our transformation leverages the distributional information of the error, we would expect that it could also improve the power over the INT method for other tests. 
We will also apply the proposed transformation to MORST and investigate the performance in simulation studies and real data analysis.   

In model~\eqref{Eq:model}, the explanatory variables $\mathbf{g}_i$ have a linear relationship with $y_i$. This is not necessary when local alternatives are considered. In fact, it is easy to see that all the asymptotic results also hold for the model
\[
y_i = \mathbf{z}_i^\top\boldsymbol{\alpha} + h(\mathbf{g}_i^\top\boldsymbol{\beta}) + \epsilon_i,
\]
where $h(\cdot)$ is a pre-specified continuously differentiable function and $|h'(0)| < +\infty$ and include more complex forms of associations.  We also investigate this more general model in our simulation studies.

\section{Simulation studies}\label{sec:simu}

\subsection{Model settings}

We conducted extensive simulations to evaluate the type I error of the proposed locally powerful kernel transformation (``LPT"), and compare its power with the rank-based inverse normal transformation (``INT") and untransformed association tests (``UAT") where no transformation is performed on the phenotype (or residual). Following the genotype data generating scheme in \cite{sun2020genetic}, we randomly selected 1,000 genes and generated the genotype $\mathbf{g}$ in every gene exhibiting linkage disequilibrium by HAPGEN2 using the CEU population from HapMap3~\citep{international2010integrating} as a reference panel.
The sample sizes $n\in \{5000, 10000\}$ were chosen to mimic large-scale genetic datasets.  For the $i$th subject, covariates were randomly generated as $\mathbf{z}_i= (1, z_{i1}, z_{i2})^\top$ with $z_{i1}\sim N(5,1)$ and $z_{i2}\sim binom(p=0.5)$.

For type I error simulations, we use the linear model $y_i =  \mathbf{z}_i^\top\boldsymbol\alpha+\epsilon_i$, where $\boldsymbol{\alpha} = (1, 0.8, 1)^\top$. The error $\epsilon_i$ follows six distributions: (1) the standard normal distribution $N(0,1)$; (2) the skew-normal distribution with location parameter 0, shape parameter 10, and scale parameter 5; (3) the chi-squared distribution with 5 degrees of freedoms $\chi^2_5$;  (4) the lognormal distribution (i.e., $\log(\epsilon_i)\sim N(0,1)$); (5) the bimodal normal distribution with 30\% probability of being in $N(0,1)$ and 70\% probability of being in $N(5,2^2)$; (6) the t distribution with 3 degrees of freedom $t(3)$. The skew-normal distribution has been considered for modeling skewed traits, such as Body Mass Index \citep{ma2005locally}. The bimodal and heavy-tailed distributions are also commonly observed in various traits, especially at the gene expression level \citep{song2017qrank,korthauer2016statistical,shalek2013single}. We do not set the error mean to be zero as a nonzero error mean can be modeled into the intercept. The kernel transformation bandwidth is selected using the \texttt{R} function \texttt{bw.nrd}.

For power analysis, we consider the location model $y_i = \mathbf{z}_i^\top\boldsymbol\alpha + \mathbf{g}_i^\top\boldsymbol{\beta} + \epsilon_i$. 
 We set $\boldsymbol{\alpha}$ to be the same as in the type I error simulation. Three factors of $\boldsymbol{\beta}$ are examined: the proportion of non-zero effects, effect signs, and effect sizes. The proportion of non-zero effects is set to be 10\%, 30\%, and 60\%, representing sparse, moderately sparse, and dense signals, respectively. Two effect signs scenarios are considered: (1) the unidirectional case, where all the nonzero $\beta_j$'s have the same sign; (2) the bidirectional case, where half of the nonzero $\beta_j$'s are positive and the other half are negative. Effect sizes are adjusted according to the error distributions and the proportion of non-zero effects to ensure an appropriate comparison of different methods. Specifically, for a 10\% proportion for non-zero effects, we set  $\beta_j = 0.3$ for $\chi_5^2$ errors, $\beta_j = 0.1$ for log-normal errors, and $\beta_j = 0.2$ for the other errors. For a 30\% proportion,  $\beta_j$ is halved. When the proportion of non-zero effects is 60\%, we set $\beta_j = 0.05$ for $\chi^2_5$ errors, $\beta_j = 0.02$ for log-normal errors, $\beta_j = 0.03$ for normal error and $t_3$ error distributions, $\beta_j = 0.05$ for skew-normal error and bimodal normal errors.

\subsection{Type I error analysis}
We examine the perseverance of nominal type I error of the proposed LPT at significant threshold  $\alpha_0\in\{0.01, 1e-5,2.5e-6\}$ with sample sizes $n\in\{5000,10000\}$, among which $\alpha_0 = 2.5e-6$ is the commonly used threshold in gene-based association analyses to account for multiple testing~\citep{auer2015rare}. We generated 10,000 Monte Carlo replicates per gene, which leads to $10^7$ replicates in total. 
The results presented in Table \ref{tab:H0} demonstrate that the Burden, SKAT, and MORST tests, when combined with our proposed transformation, maintain the nominal type I error across all levels.   This is due to the large sample sizes used in the simulation. In biobank-scale studies, where the sample size $n$ is typically much larger than 10,000, the type I error would also be well-controlled.

\begin{table}[!ht]
\centering
\begin{tabular}{ll|lll|lll}
\toprule
error & Sig.thres & \multicolumn{3}{c|}{\underline{$n=5000$}}& \multicolumn{3}{c}{\underline{$n=10000$}}\\
&& Burden & SKAT &  MORST& Burden & SKAT &  MORST \\ 
  \hline
 \underline{$N(0,1)$}& 1e-02 &  9.97e-03 & 9.99e-03 &  9.91e-03& 1.00e-02 & 1.00e-02 & 9.94e-03  \\ 
 & 1e-05 &  1.04e-05 & 9.92e-06 &  9.82e-06 & 9.80e-06 & 9.70e-06 & 1.07e-05\\ 
 & 2.5e-06 & 2.61e-06 & 2.30e-06 &  2.50e-06  & 1.90e-06 & 2.30e-06 & 3.00e-06 \\ 
 
  \hline
  \underline{binorm}&   1e-02 & 1.00e-02 & 9.93e-03 &9.85e-03& 9.99e-03  &9.96e-03    &9.93e-03   \\
&  1e-05 &  8.07e-06 & 8.42e-06 & 7.36e-06 & 1.03e-05 & 1.01e-05  & 1.02e-05 \\  
&  2.5e-06 & 2.11e-06 & 2.01e-06  & 2.22e-06  &2.77e-06 & 3.06e-06 & 1.92e-06 \\ 
    
    \hline
     \underline{$t_3$}& 1e-02 &  9.98e-03 & 9.98e-03  & 9.92e-03 & 9.96e-03  &1.00e-02   &9.92e-03   \\  
&  1e-05 &  9.80e-06 & 1.05e-05  & 1.06e-05& 9.08e-06  & 1.00e-06 & 9.87e-06\\ 
& 2.5e-06 &2.30e-06 &  2.49e-06 & 3.16e-06  &  1.92e-06  &  2.58e-06 & 2.87e-06   \\ 
     \hline
 \underline{sknorm}& 1e-02 &  9.94e-03 & 9.94e-03 &   9.94e-03& 1.00e-02 & 9.98e-03 & 1.00e-02  \\ 
&1e-05 & 9.72e-06 & 9.52e-06 &  1.09e-05 & 9.21e-06 & 1.03e-05 & 8.38e-06\\ 
 & 2.5e-06  & 2.43e-06 & 2.94e-06 & 3.34e-06 &   2.17e-06 & 1.97e-06 & 1.96e-06 \\ 
   \hline
   \underline{$\chi_5^2$}&  1e-02 & 1.00e-02 & 9.97e-03 &  9.94e-03& 1.00e-02 & 9.99e-03& 9.97e-03 \\
   &  1e-05 &  9.30e-06 & 9.30e-06 & 8.40e-06& 9.40e-06 & 8.90e-06  & 8.80e-06\\ 
   &2.5e-06 &  2.20e-06 & 2.20e-06 &  1.80e-06 & 2.10e-06 & 2.00e-06& 1.70e-06 \\ 
   \hline
  \underline{lognorm}& 1e-02 & 1.00e-02 & 9.95e-03 &9.95e-03& 1.00e-02 & 1.00e-02  & 1.00e-02 \\ 
  & 1e-05 &  1.06e-05 & 1.10e-05 & 9.90e-06 & 1.20e-05 & 1.05e-05  & 1.00e-05 \\ 
  &2.5e-06 & 2.70e-06 & 2.40e-06  & 2.00e-06  & 2.90e-06 & 3.40e-06  & 2.50e-06\\  
\bottomrule
\end{tabular}

\caption{Simulation results for type I error analysis based on $10,000,000$ replicates.}\label{tab:H0}
\end{table}

\subsection{Power analysis}
 To evaluate the power of the proposed method, we generated 10 Monte Carlo replicates per gene, resulting in $10^4$ replicates in total. The significance level is $2.5e-6$. 
 We present the results for $n=10,000$ in Figures \ref{fig:power_1}-\ref{fig:power_2} and the results for $n=5,000$ in Supplement Section 2.
The results across all effect settings and test statistics show that the proposed LPT consistently outperforms INT and UAT under non-Gaussian errors, while also being as powerful as these methods under Gaussian errors. This is expected since our LPT leverages error density information to enhance power. Notably, LPT exhibits substantial power improvement under highly skewed errors (e.g., lognormal and skewed normal) and bimodal normal errors.
We note that the simulation results from \cite{mccaw2020operating} indicate that INT is significantly more powerful than UAT under lognormal errors, aligning with our findings for these two methods.  Moreover, the proposed LPT further enhances the performance of INT under lognormal errors. Comparisons among the three types of test statistics—Burden, SKAT, and MORST—are consistent with those observed in Gaussian linear regression settings \citep{wu2011rare,liu2022minimax}.
The results with $n=5,000$ lead to the same conclusion (Supplement Section 2).

\begin{figure}[!ht]
    \centering
   
    \includegraphics[scale = 0.355]{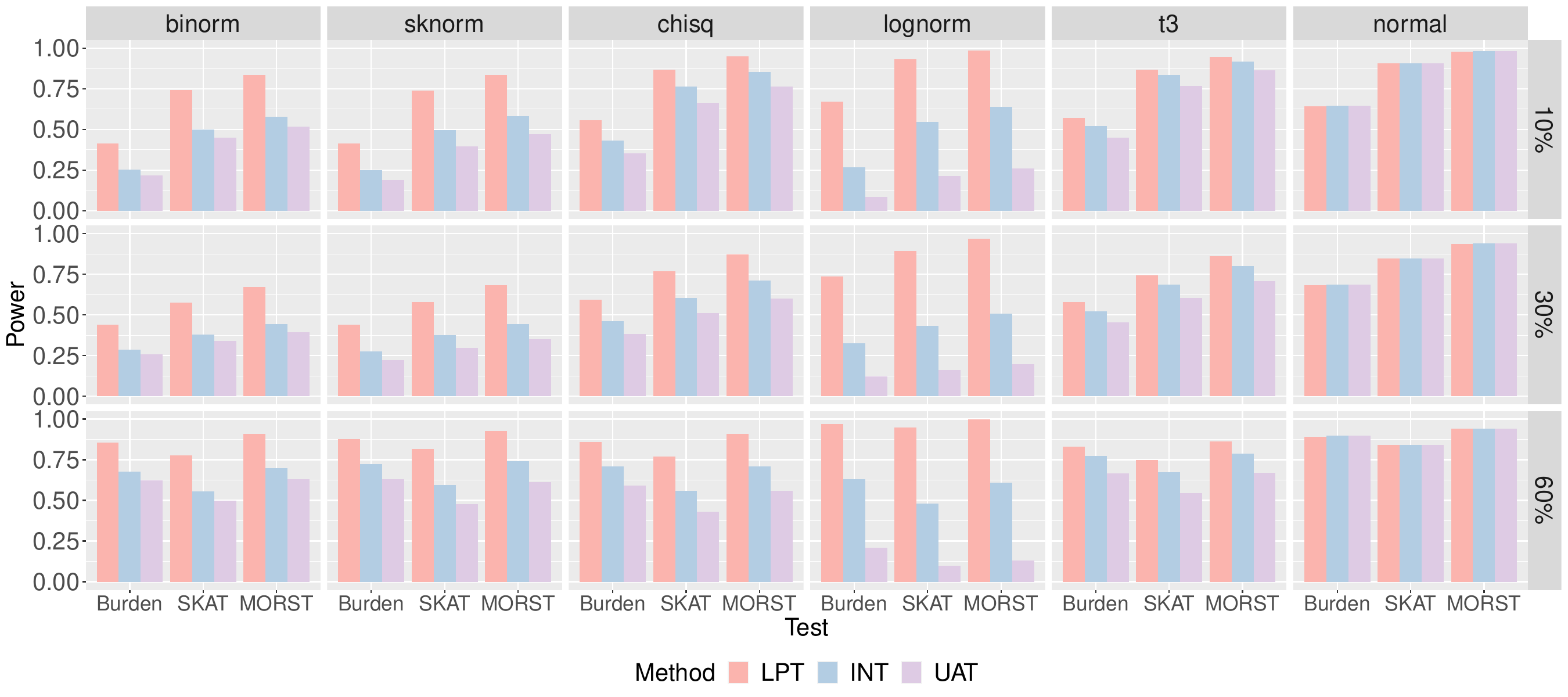}
\caption{ Power comparisons of the three transformation methods under the Burden, SKAT, and MORST tests. The columns correspond to the six different error distributions, respectively. The three rows correspond to the three proportions of non-zero effects. The effects are \emph{unidirectional}, and the sample size $n$ is 10,000. 
}
    \label{fig:power_1}
\end{figure}
\begin{figure}[!ht]
    \centering
  
   \includegraphics[scale = 0.355]{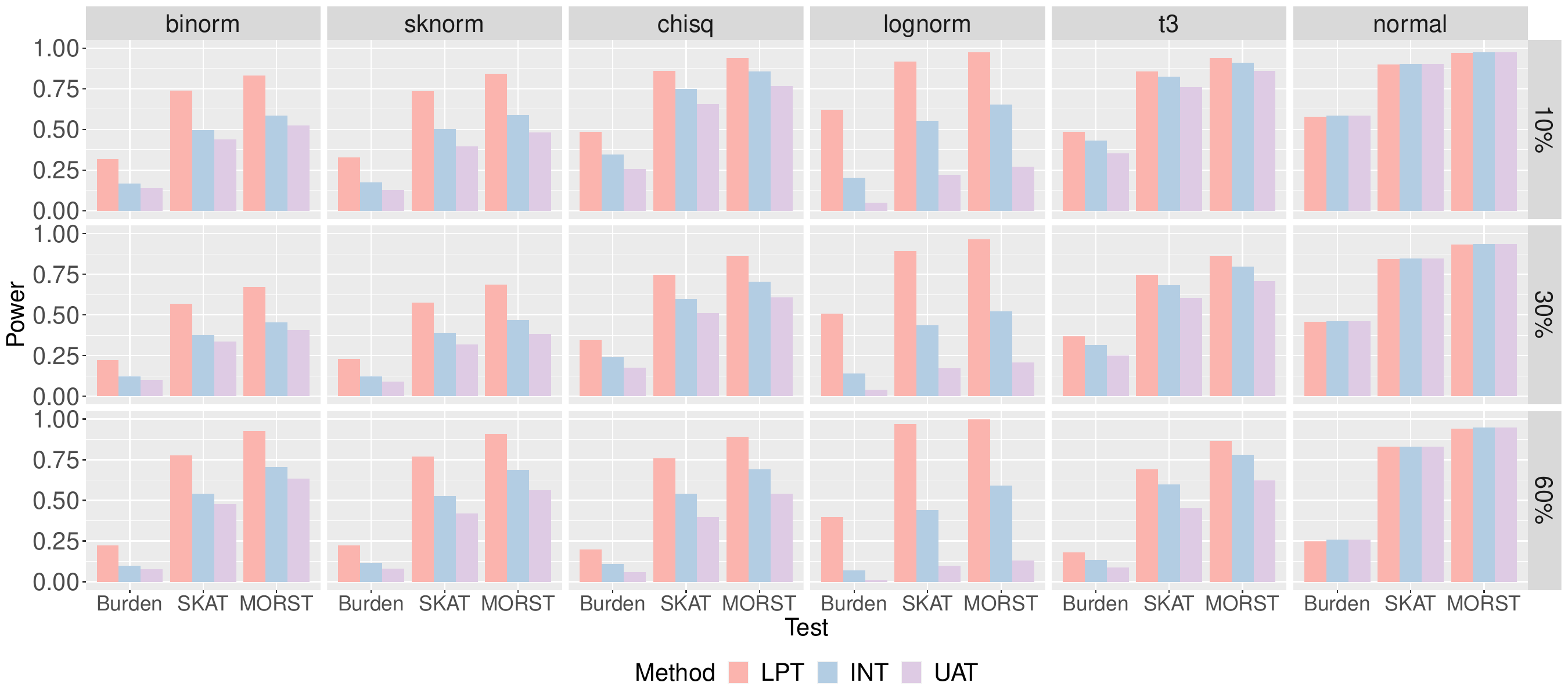}
\caption{Power comparisons of the three transformation methods under the Burden, SKAT, and MORST tests. The columns correspond to the six different error distributions, respectively. The three rows correspond to the three proportions of non-zero effects. The effects are \emph{bidirectional}, and the sample size $n$ is 10,000. }
    \label{fig:power_2}
\end{figure}

\subsection{Computation time}\label{Secsub:comput:time}

We demonstrate the computation time of the proposed transformation method based on our implementation. Note that the transformation of phenotypes or residuals is just one step in gene-based (or variant set) analysis. Other steps include fitting the null model and performing the statistical tests.  The computation time here is only about the transformation.
We performed the computations on a desktop with 2.5 GHz Intel i7-10700F CPU and 16 GB memory. Since the computation time does not depend on the distribution of $Y$, we simulate $Y$ from the standard normal distribution and consider a wide range of sample sizes: $n = 1\times 10^4, 2\times 10^4, 5\times 10^4, 1\times 10^5 ,2\times 10^5, 5\times 10^5$. Table~\ref{tab:comput_time} shows the computation time in seconds at different sample sizes.

The computational burden becomes substantial when the sample size is at the biobank-scale (e.g., $n = 2\times 10^5$ or $5\times 10^5$). It is feasible to perform the proposed transformation method once in gene-based analysis for large $n$. However, it would be computationally impractical if the transformation had to be carried out for every gene. As described in Section~\ref{sec:est}, our method only requires performing the transformation once because the residuals under the null are used in the estimation of the transformation function.

\begin{table}
    \centering
    \begin{tabular}{c|cccccc}
        \toprule
        $n$ & $1\times 10^4$ & $2\times 10^4$ & $5\times 10^4$ & $1\times 10^5$ &  $2\times 10^5$ & $5\times 10^5$  \\ \hline
        Computation time & 1.17 & 4.37 & 24.50 & 90.31 & 328.89 &  1801.04 \\ \bottomrule
    \end{tabular}
    \caption{Computation time in seconds of performing the proposed transformation once}
    \label{tab:comput_time}
\end{table}

\subsection{Additional simulation studies}
\begin{figure}[!ht]
    \centering
    \includegraphics[scale = 0.35]{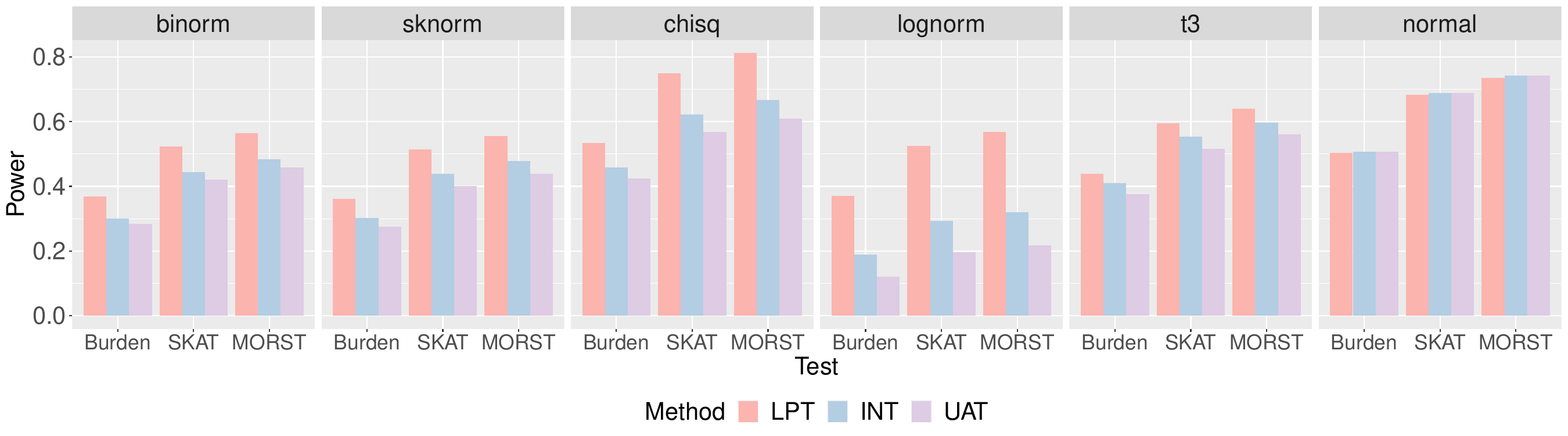}
\caption{Power comparisons of the three transformation methods under the Burden, SKAT, and MORST tests. The columns correspond to the six different error distributions, respectively. The three rows correspond to the three proportions of non-zero effects. Data is generated from the \emph{quadratic model} with 10\% non-zero and unidirectional effects. The sample size $n$ is 10,000. }
    \label{fig:power_4}
\end{figure}

We further evaluate the power of LPT under the quadratic model, which has also been studied for nonlinear genetic effects, such as variance quantitative trait locus \citep{lin2022accounting}. We generate $y_i$'s from the quadratic model with $y_i = (\boldsymbol{g}_i^\top\boldsymbol{\beta})^2 + \boldsymbol{z}_i^\top\boldsymbol{\alpha} + \epsilon_i$. It is an example of Section~\ref{sec:ext} with $h(\cdot)$ being a quadratic function. Results suggest that LPT also maintains a power improvement over the other two methods (Figure~\ref{fig:power_4}). This demonstrates that all the asymptotic results hold and LPT is locally optimal when the $y_i$ is associated with $\mathbf{g}_i$ through the general form of $h(\mathbf{g}_i^\top\boldsymbol{\beta})$. 

\section{Application on UK Biobank data}\label{sec:data}

\subsection{Data introduction and analysis results}

\begin{figure}[!ht]
    \centering
    \includegraphics[scale = 0.5]{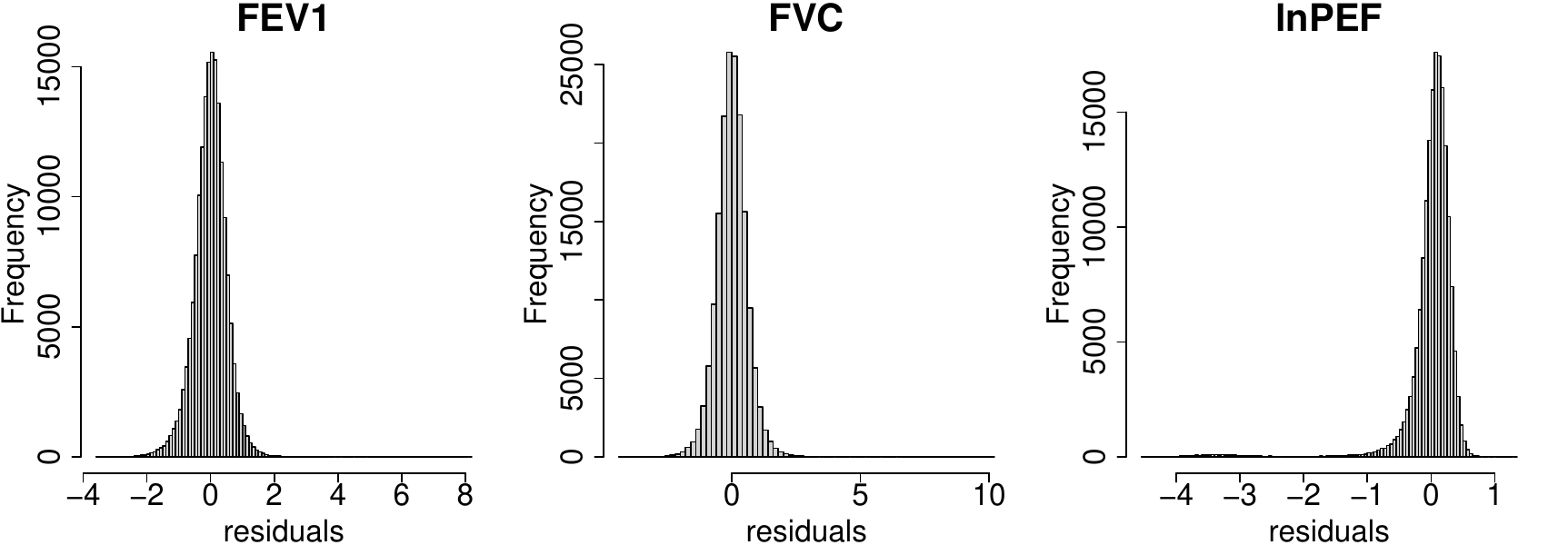}
    
    \includegraphics[scale = 0.5]{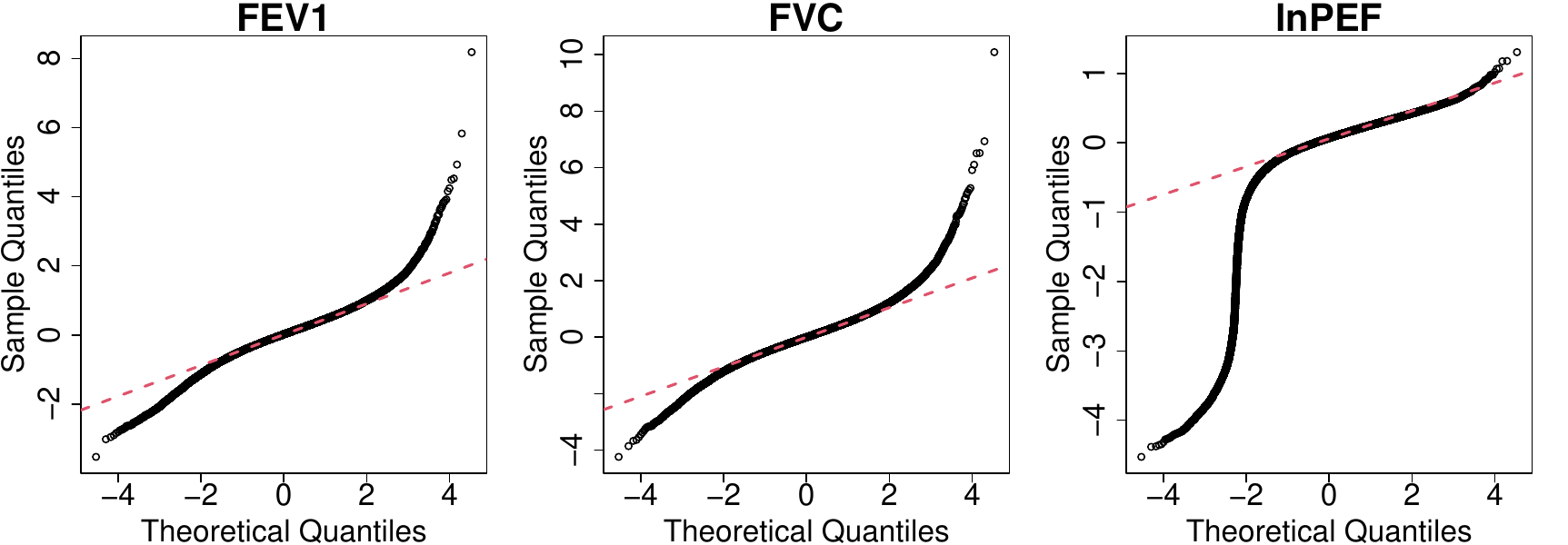}
    \caption{Residual examination with histogram (Top) and normal Quantile-Quantile plot (Bottom) for three traits, respectively. }
    \label{fig:res_plot}
\end{figure}

To demonstrate the proposed transformation in real data, we apply all methods to independent subjects from the British White group in UK Biobank data (https://www.ukbiobank.ac.uk/). We perform gene-based analysis for three spirometry measurement traits: Forced Expiratory Volume in one second (FEV1), Forced Vital Capacity (FVC), and the logarithm of Peak Expiratory Flow (lnPEF). These measurements are crucial in respiratory medicine for diagnosing, monitoring, and managing various lung diseases. The adjusted covariates include age, sex, BMI, height, and the top 20 genetic PCs as in \cite{mccaw2020operating}. These three traits have highly skewed and asymmetric residual distributions (Figure \ref{fig:res_plot}). After removing individuals with missing values of the above traits and covariates, a total of 221,257 subjects remained. We use 171,257 individuals as the main set, while the remaining 50,000 individuals are considered as a validation set. We restrict our analysis to SNPs with MAF $\geq 0.05$ and only consider genes with more than 5 SNPs in the defined range. For each gene, we perform three set-based association tests (i.e., Burden, SKAT, and MORST), and each test is accompanied by the three transformation methods (i.e., LPT, INT, and UAT). The results are based on the genome-wide significant threshold of $2.5e-6$ that is commonly used for gene-based association studies~\citep{auer2015rare}.

The number of genes identified by each method in the main set ($n=171,257$) are presented in Table \ref{tab:ukb_17w_nokinship}. 
The proposed LPT identified more significant genes than both INT and UAT across all three traits and tests. 
Notably, there is a pronounced improvement in the number of genes identified by LPT for the trait lnPEF, which exhibits substantial skewness in its residual distribution as shown by the QQ plot in Figure \ref{fig:res_plot} (bottom row). These results for lnPEF are consistent with our simulations under highly skewed and asymmetric errors, demonstrating considerable power improvements from LPT compared to INT. Additionally, the real data analysis in \cite{mccaw2020operating} shows that INT detected significantly more signals than UAT for the trait lnPEF, indicating that the transformation methods could have dramatically different performances when the error distribution is substantially skewed and asymmetric.
Within each transformation method, MORST is more powerful than SKAT and Burden, which is consistent with the literature~\citep{liu2022minimax}.

\begin{table}[ht]
\centering
\begin{tabular}{r|rrr|rrr|rrr}
  \hline
&\multicolumn{3}{c|}{\underline{Burden}} &\multicolumn{3}{c|}{\underline{SKAT}}&\multicolumn{3}{c}{\underline{MORST}}\\
 & LPT & INT & UAT&LPT & INT & UAT&LPT & INT & UAT \\
  \hline
  
FEV1 & 77 & 63 & 64 & 156 & 129 & 129 & 229 & 203 & 204 \\ 
  FVC & 59 & 50 & 52 & 122 & 99 & 96 & 200 & 161 & 161\\ 
  lnPEF & 81 & 51 & 2 & 181 & 119 & 13& 217 & 165 & 50 \\ 

   \hline
    
\end{tabular}
\caption{Number of genes identified based on 171,257 individuals from UK Biobank. }\label{tab:ukb_17w_nokinship}
\end{table}

\subsection{Discovery gain and validation analysis}
We define the empirical discovery gains between LPT and INT as in~\cite{mccaw2020operating}:
\[
\text{Discovery Gain }_{(\text{LPT\ vs\ INT})} = \frac{n_{\text{LPT}}-n_{\text{INT}}}{n_{\text{LPT}\cup \text{INT}}},\quad \text{Discovery Gain }_{(\text{INT\ vs\ LPT})} = \frac{n_{\text{INT}}-n_{\text{LPT}}}{n_{\text{LPT}\cup \text{INT}}}.
\]
A larger discovery gain indicates more unique genes identified by the first transformation in the numerator.
Within three tests, namely Burden, SKAT, and MORST, we compare the discovery gain between LPT and INT. We observe that LPT markedly identifies more unique genes than INT across all cases (Table \ref{tab:DG}), showcasing its exceptional performance. 

\begin{table}[ht]
 
\centering
\begin{tabular}{r|rr|rr|rr}
  \hline
  &\multicolumn{2}{c|}{\underline{Burden}}&\multicolumn{2}{c|}{\underline{SKAT}}&\multicolumn{2}{c}{\underline{MORST}}\\
 & LPT vs INT & INT vs LPT  & LPT vs INT & INT vs LPT & LPT vs INT & INT vs LPT \vspace{-1mm}\\ 
  \hline
FEV1 & 20.25\% & 2.53\%  & 21.34\% & 4.88\% & 16.80\% & 6.15\% \\ 
  FVC & 20.63\% & 6.35\%  & 22.66\% & 4.69\% & 23.70\% & 5.21\% \\ 
  lnPEF & 37.04\% & 0.00\% & 34.25\% & 0.00\% & 25.00\% & 1.36\% \\ 
   \hline
\end{tabular}

\caption{Results on discovery gain based on genes identified from 171,257 individuals.}
\label{tab:DG}
\end{table}

As a validation of our discoveries with 171,257 individuals, we further conducted the same analysis using the validation set with 50,000 individuals, independent from the main set. We consider a gene as ``reproducible" if its $p$-value is smaller than 0.05 in the validation set while smaller than $2.5e-6$ in the main set. The percentage of reproducible genes identified in Table \ref{tab:ukb_17w_nokinship} is reported in Table \ref{tab:ukb_5w_nokinship}.  We observed a high proportion of reproducible genes for all methods (Table \ref{tab:ukb_5w_nokinship}), indicating the validity of genes identified in the main set. The small portion of genes that are not successfully reproduced could be related to their weak signals, such that 50,000 individuals may not be sufficient to claim their significance. 

\begin{table}[!ht]

\centering
\begin{tabular}{r|rrr|rrr|rrr}
  \hline
&\multicolumn{3}{c|}{\underline{Burden}} &\multicolumn{3}{c|}{\underline{SKAT}}&\multicolumn{3}{c}{\underline{MORST}}\\
 & LPT & INT & UAT&LPT & INT & UAT&LPT & INT & UAT\\
  \hline
  
FEV1 & 62 &  53 &  54 & 136 & 109 & 111 & 187 & 157 & 158 \\ 
&80.52\% & 84.13\% & 84.38\% & 87.18\% & 84.50\% & 86.05\% & 81.66\% & 77.34\% & 77.45\% \\ 
\hline
  FVC &  50 &  41 &  42 &  96 &  74 &  73 & 133 & 106 &  97 \\ 
  & 84.75\% & 82.00\% & 80.77\% & 78.69\% & 74.75\% & 76.04\% & 66.50\% & 65.84\% & 60.25\% \\  \hline
  lnPEF &  69 &  41 &   2 & 164 &  99 &  13 & 186 & 138 &  47 \\ 
  & 85.19\% & 80.39\% & 100.00\% & 90.61\% & 83.19\% & 100.00\% & 85.71\% & 83.64\% & 94.00\% \\ 

   \hline
    
\end{tabular}

\caption{Results on reproducibility using 50,000 independent subjects from UK Biobank. For each trait, the two rows denote the number and the portion of reproducible significant genes ($p<0.05$ in the validation set and $p<2.5e-6$ in the main set), respectively.}\label{tab:ukb_5w_nokinship}
\end{table}

We further visualize the Quantile-Quantile plot for LPT against INT across traits and tests (Supplement Figure 4). In all scenarios, we observe that the two methods align well when the $p$-values are large. As INT is shown to control type I errors at the nominal levels~\citep{mccaw2020operating}, this suggests that LPT also maintains the controlled type I errors in the application. The deviation in the trait InPEF, which is evident when the p-value becomes smaller, indicates that LPT will be more powerful than INT as we discussed in previous results (Table~\ref{tab:ukb_17w_nokinship}).

Finally, we perform a pooled analysis with all 221,257 individuals and present the results in Table \ref{tab:results_22w}. With a larger sample size, all methods identified more genes, but the exemplary performance of LPT was maintained and continued dominating INT and UAT. 

\begin{table}[ht]
\centering

\begin{tabular}{r|rrr|rrr|rrr}
  \hline
  &\multicolumn{3}{c|}{\underline{Burden}} &\multicolumn{3}{c|}{\underline{SKAT}}&\multicolumn{3}{c}{\underline{MORST}}\\
 & LPT & INT & UAT&LPT & INT & UAT&LPT & INT & UAT \\
  \hline
  
FEV1 & 112 & 92 & 90 & 229 & 212 & 204 & 338 & 291 & 286\\ 
  FVC & 87 & 79 & 73 & 188 & 167 & 161 & 289 & 257 & 252\\ 
  lnPEF & 119 & 76 & 15 & 232 & 162 & 41& 308 & 212 & 91 \\ 
   \hline
\end{tabular}

\caption{Identified genes based on all 221, 257 individuals. }\label{tab:results_22w}
\end{table}

\subsection{Practical insights}
 Several genes uniquely identified using LPT have been validated in the literature. For example, HLX and AFF1 were identified as significant genes shared between GlycA (a systemic inflammatory marker) and lung function through transcriptome-wide association studies \citep{guo2024genetic}. HLX has been implicated in both FEV1 and FVC lung function parameters, suggesting a role in regulating pulmonary capacity, potentially through immune response pathways. AFF1, a member of the AF4/lymphoid nuclear protein family, was identified as sharing genetic signals between GlycA and FEV1. 
Additionally, EML4 was identified as a gene with important connections to lung function, particularly through its association with the FEV1/FVC ratio \citep{kachuri2020immune}. TGFB2, another gene uniquely identified by our method, has significant associations with multiple lung function parameters, including FEV1/FVC, FEV1, FVC, and PEF \citep{shrine2023multi}. Notably, TGFB2 is supported by multiple variant-to-gene criteria, providing strong evidence for its causal role in lung function regulation.

In our analysis of the UKB data, transformation of the phenotype for 221,257 individuals only needs to be carried out once and takes approximately 5.46 minutes on a desktop with 2.5 GHz Intel i7-10700F CPU and 16 GB memory. In contrast, the gene-based association testing across the genome requires tens of CPU hours for the Burden test, and hundreds of hours for SKAT and MORST. The one-time transformation step incurs negligible computational cost relative to the extensive runtime of the association tests. 
Through parallel execution using, e.g., 16 computing cores, the whole association analysis can be done within one day.

\section{Discussion}

Motivated by the need for biobank-scale studies, we propose a computationally scalable transformation method to improve the detection of weak signals. For linear-type and SKAT-type tests, we show that the proposed transformation leads to locally most powerful tests and, therefore, is particularly powerful against alternatives with weak signals. Asymptotic null distributions of the test statistics with transformed residuals are established. A computationally efficient estimator for the transformation is also proposed and is shown to be consistent and lead to optimal local power under weak signals. The analysis of UK biobank data and simulation studies demonstrate that our method controls the type I error well under a moderately large sample size and has a higher power than the INT method. 

The rationale of our new transformation method centers around the massive size, an important feature in biobank-scale studies. The massive sample size alleviates the concern of type I error inflation based on asymptotic null distributions and, meanwhile, leads to the interest in detecting weak signals. On the other hand, the massive sample size increases the computational burden and requires scalable statistical methods. We take advantage of weak signals to deal with the computational complexity, and propose a consistent estimator of the transformation that is computationally scalable for gene-based analysis.

In genome-wide association studies, mixed models are commonly used to account for sample relatedness or confounding bias, which violates the assumption of independent error terms ($\varepsilon_i$'s). Extending our method to derive an optimal transformation under sample relatedness is not straightforward, as it requires specifying the likelihood function. A comprehensive treatment of sample relatedness is beyond the scope of this work and warrants further investigation. Here, we offer some considerations and a potential approach for addressing this issue. Recall that the transformation function affects only the power but not the asymptotic null distribution. This implies that a suboptimal transformation could still control type I errors but may not maximize power. Moreover, in large-scale studies such as the UK Biobank, the genetic relatedness matrix is often sparse. Statistically, this implies that the overall dependency among samples is weak and does not significantly deviate from independence. Therefore, a possible modification of our method to address sample relatedness proceeds as follows: First, select a subset of independent samples and use them to derive the transformation function based on our method. Next, apply this transformation to the phenotypes of all samples and perform association tests using mixed models. While this transformation may not be optimal under sample relatedness, it could still improve power relative to other transformations, such as INT, with controlled type I errors.
    
Our proposed transformation is developed under the model~\eqref{Eq:model} that assumes additive effects and the homogeneity of variance. It has been observed that the variance of quantitative traits could differ with the genotype, especially when the trait distribution is non-Gaussian~\citep{young2018identifying}. This phenomenon is caused by the existence of interaction effects among SNPs~\citep{struchalin2010variance}. A possible future direction is to improve the existing approach and fine-tune in the presence of non-constant variances.

\section*{Acknowledgements}

This research has been conducted using the UK Biobank Resource under Application Numbers 154421 and 207159.
Y.L.’s research was supported by
the National Natural Science Foundation of China (Nos. 12371282 and 72495122). The authors thank the Associate Editor and two referees for insightful and constructive comments, which have greatly helped improve the paper.

\section*{Supplementary materials}
Supplementary materials include (A) a supplement that provides the proofs of theorems~\ref{Thm:optimal:test}--\ref{Thm:kernel} and other technical lemmas, additional simulation results, as well as additional results for the UKB data analysis; (B) reproducibility materials that provide the R code to reproduce all the results presented in the paper.

%\clearpage

\bibliographystyle{apalike}
\bibliography{paper-ref}

\end{document}